\documentclass[11pt,preprint]{aastex}

\slugcomment{This version 1 April 2005}

\newcommand{\faa}{20~cm}
\newcommand{\fluxa}{S$_{20 \rm \thinspace cm}$}
\newcommand{\fbb}{11~cm}
\newcommand{\fluxb}{S$_{11 \rm \thinspace cm}$}
\newcommand{\fcc}{6~cm}
\newcommand{\fluxc}{S$_{6 \rm \thinspace cm}$}
\newcommand{\fdd}{3~cm}
\newcommand{\fluxd}{S$_{3 \rm \thinspace cm}$}

\newcommand{\mj}{~$\mu$Jy}

\newcommand{\myr}{$M_{\sun}yr^{-1}$}

\shorttitle{ATHDFS survey: Description and Initial Results}
\shortauthors{Norris et al.}

\begin{document}

\title{Radio Observations of the Hubble Deep Field South
region: \\
    I. Survey Description and Initial Results}

\author{Ray P. Norris}
\affil{CSIRO Australia Telescope National Facility, PO Box 76, Epping
NSW 1710, Australia. Ray.Norris@csiro.au}

\author{Minh T. Huynh}
\affil{Research School of Astronomy \& Astrophysics, Mount Stromlo
Observatory, The Australian National University, Canberra ACT 2611,
Australia}

\author{Carole A. Jackson, Brian J. Boyle, Ronald. D. Ekers, Daniel A.
Mitchell, Robert J. Sault, Mark H. Wieringa}
\affil{CSIRO Australia Telescope National Facility, PO Box 76, Epping
NSW
1710, Australia}

\author{Robert E. Williams}
\affil{Space Telescope Science Institute 3700, San Martin Drive,
Baltimore, MD
21218, USA}

\author{Andrew M. Hopkins\altaffilmark{1}}
\affil{Department of Physics \& Astronomy, University of Pittsburgh,
3941
O'Hara Street, Pittsburgh, PA15260, USA}

\and

\author{James Higdon}

\affil{Department of Astronomy, 
Cornell University,
215, Space Sciences Building,
Ithaca, NY 14853-6801, USA}

\altaffiltext{1}{Hubble Fellow}

 
\begin{abstract}
This paper is the first of a series describing the results of the
Australia Telescope Hubble Deep Field South (ATHDFS) radio survey. The survey was conducted at four wavelengths – 20, 11, 6, and 3 cm, over a 4-year period, and achieves an rms sensitivity of about 10 \mj\ at
each wavelength. We describe the observations and data reduction
processes, and present data on radio sources close to the centre of the
HDF-S. We discuss in detail the properties of a subset of these sources.
The sources include both starburst galaxies and galaxies powered by an
active galactic nucleus, and range in redshift from 0.1 to 2.2. Some of
them are characterised by unusually high radio-to-optical luminosities,
presumably caused by dust extinction.

\end{abstract}


\section{Introduction}

Following the success of the original Hubble Deep Field project in the
northern sky (HDF-N; \citet{wil96}) the WFPC2, STIS and NICMOS
instruments of the Hubble Space Telescope (HST) were used to produce
very deep images of a corresponding southern sky region - the `Hubble
Deep Field South' (HDF-S; Williams et al. 2000). The HDF-S differs from the
HDF-N in that a field was chosen near a bright, high redshift (z=2.24)
quasar to enable a study of the intervening intergalactic medium
\citep{out99}.  

Both Hubble Deep Fields have provided a focus for a wide range of
research, and extensive ground-based observations have complemented the
optical and infrared images obtained by the HST, leading to significant
advances in our understanding of the evolution of the Universe. However,
most of these observations were conducted at optical and infrared  wavelengths. 

Radio observations occupy an important observational niche as they probe
physical conditions that are faint or unobservable at optical/infrared
wavelengths, and they are also unaffected by dust extinction. The HDF-N
has been studied in detail at radio wavelengths using deep Very Large
Array (VLA)  observations \citep{ric98,ric00} resulting in the most sensitive radio images to date, reaching an rms
sensitivity of 4~\mj\ at \faa, and 1.6~\mj\ at \fdd. 

At high flux density levels, radio source surveys are generally
dominated by radio-loud active galaxies and quasars. However, at \mj\
levels, the source counts are increasingly dominated by a population of
starburst galaxies \citep{hop99}, which are generally believed to be
high-redshift (and in some cases higher-luminosity) counterparts to the
ultra-luminous infrared galaxies typified by Arp220. \citet{mux00} have shown that, in their sample,  $\sim$70\% of the \mj\ sources
are starburst-type sources lying at redshifts between z $\sim$ 0.4 and
1.0. Another $\sim$20\%  are low-luminosity AGNs in elliptical galaxies
at  z $\sim$ 1. The remaining 10\% of \mj\ sources have
optically-faint hosts, close to or beyond the HDF-N limit. 

Further support for both starburst and AGN processes, based on observations at radio,  far-infrared
(FIR) and submillimetre (submm) wavebands, have been given by  \citet{bar00}, \citet{afo01},
\citet{bru02}, \citet{cha03}, \citet{ber04}, and \citet{geo04}.
These observations suggest that there is a large
population of highly-obscured, but very active galaxies at z\,=\,1--5
which is responsible for producing the bulk of the extragalactic
background in the FIR/submm region. These galaxies may host the major sites of
massive star formation and fuelling of active galactic nuclei (AGN) at
high redshifts. Moreover, there is tentative evidence that
these galaxies are strongly clustered, with a correlation-length
exceeding any other known high-z population \citep{bla04, ste03}. These
results hint at a direct link between these dusty, high-redshift
galaxies and the growth of large scale structure in the early Universe,
which may be the origin of the basic environmental variations of galaxy
properties (e.g. morphologies, stellar ages) seen in the modern
Universe.

Our understanding of these phenomena requires  sensitive radio
observations, extending to the limits of current instrumentation,
together with equally sensitive observations at other wavelengths. The
combination of deep radio observations, deep
HST observations, and deep observations at other wavelengths, provides a
remarkable opportunity in the Hubble Deep Fields to study the evolution of these galaxies.
This is the first of a series of papers presenting results from our
deep radio survey of the HDF-S made with the Australia
Telescope Compact Array (ATCA). The principal goals of this survey are
to: 

\begin {itemize}
\item { Use source counts together with redshifts to study the
cosmological evolution of the star formation rate, the evolutionary
history of galaxies, and  the origin of current stellar populations.}

\item {Understand the  evolutionary relationship between galaxies and
AGN and trace the development of supermassive black holes.}

\item {Trace the changing balance between obscured and unobscured
activity over the lifetime of the Universe. Obscured activity produces
no more than 30\% of the total bolometric luminosity in the local
Universe, whereas it has been suggested that high-redshift  obscured
activity might account for 90\% of the 
detected luminosity. We also need to relate this transformation to the
changing roles of star formation and AGN activity.}

\end {itemize}

Paper 1 (this paper) describes the observations and data reduction,
presents the radio data at all four wavelengths in the inner  region of
the HDF-S, and discusses a representative subset of sources in this
region in some detail.

Paper 2 \citep{P2} presents the extended 20 cm observations and
catalogue of detected sources, and analyses the differential radio
source counts.

Paper 3 \citep{P3} presents the extended 13, 6, and 3 cm observations,
and discusses radio spectral indices and individual radio sources in the
field.

Paper 4 \citep{P4} presents optical identifications, classification, and
redshifts, and discusses the implications for the radio luminosity
function, and the cosmological evolution of the star formation rate.

Throughout this paper, we use the following cosmological parameters:
$H_{0}=71 kms^{-1}Mpc^{-1}$, $\Omega_{m}$=0.27 and
$\Omega_{\Lambda}$=0.73.

\section{Observations}

\subsection {Observations of HDF-S Candidate Fields}

As part of the field selection process for the HDF-S, prior to the HST
observations, we observed candidate fields with the ATCA  to ensure that
the selected field had no strong radio sources that would prevent deep
radio imaging to complement the deep optical image. At the same time,
the candidate fields were searched by other groups to find a
high-redshift quasar, and were studied at other wavelengths to search
for extinction or infrared cirrus.

In May 1997, we made ATCA observations of eight candidate fields at 21
cm. All fields were observed in a single 12-hour run in a snapshot
observing mode. Each image covered a 68-arcmin square field, with the
central 34  arcmin square being the primary candidate area. Table 1
shows the field centres and the resulting source statistics. 

Our results showed that  the only detected quasar (which we identify
below as \\
ATHDFS\_J223337.5-603329 ) was only 10 arcmin from a strong radio source
(identified in Paper 2 as
ATHDFS\_J223355.6$-$604315), presenting the risk that dynamic range
limitations might reduce the sensitivity available at radio wavelengths.
However, it was judged that this source would not degrade the radio
observations sufficiently to justify moving the HDF-S to another
candidate field, provided the precise location of the HDF-S was chosen
to avoid the radio source whilst keeping the quasar within the field.
The position of the HDF-S was selected accordingly, and the results
presented here confirm the wisdom of that decision.

\subsection {Survey Observations}
We imaged the HDF-S using the ATCA at four wavelengths: 20, 11, 6 and 3 cm,  obtaining resolutions of 6, 4, 2 and 1 arcsec
respectively. Observations were taken over a period of four years from
mid 1998 to late 2001, in a variety of configurations to maximise {\em
uv} coverage (i.e. sampling of the Fourier plane). The shortest baseline was
31 m, and the longest 6000 m. Table 2 summarises the observations, and Figure 1 shows
the pointing centres. A single pointing centre was used at each wavelength, rather than a mosaic, to maximise the sensitivity that might be obtained in the available observing time.

The pointing centres differ at the four frequencies because of the
different primary beam size at the four wavelengths, and thus the
different strategies necessary to deal with the strong confusing source
ATHDFS\_J223355.6$-$604315 with \fluxa = 155 mJy. At 6 and \fdd\ the
observations are centred on the WFPC2 field, so that the strong source
is well outside the primary beam. At \faa\ and \fbb\ the pointing centre (RA =
22h 33m 37.6s and Dec = $-$60$^\circ$ 33' 29'' (J2000) ) was chosen to
be approximately half-way between the WFPC2 field  and
ATHDFS\_J223355.6$-$604315. This ensures that the strong source is well
inside the primary beam to minimise calibration problems.

Throughout the observations, the correlator was used in continuum mode 
($2 \times 128$ MHz bandwidth), with each 128 MHz band divided into $32
\times 4$ MHz channels, and all four Stokes parameters were measured. This correlator configuration achieves the highest  sensitivity, but at the price of higher resolution spectral
information. All observations were made with two 128 MHz bands. In some cases, these two bands were assigned to adjacent observing frequencies (e.g. 1344 and 1472 MHz), and in others, they were assigned to widely separated bands (e.g. 1384 and 2368 MHz) to minimise contamination by interference. 
In Table~\ref{tab2} the total hours
column is the total amount of time spent on each 128 MHz band, and
should be halved to give an equivalent observing time with two 128 MHz
bands.

The primary flux density calibrator used was PKS B1934-638, which is the
standard calibrator for ATCA observations ($S = 14.95$ Jy at 1.380 GHz;
\cite{rey94} ). We calibrated the complex
antenna gains by frequently (typically every 20-40 minutes, depending upon atmospheric phase stability) observing the secondary calibrators PKS
B2205-636 and PKS B2333-528. The resulting phase errors are typically at the level of a few degrees before self-calibration, and are not a significant limiting factor in the resulting images.

\section{Data Reduction}

\subsection {Imaging}
We used the Australia Telescope National Facility (ATNF) release of the
{\em MIRIAD} \citep{sault95} software to reduce our data, and based our imaging and source extraction process on the procedure outlined by \citet{pra00}.  

Because of the
large observing bandwidth ($2 \times 128$ MHz), the multi-frequency
synthesis \citep{sault94}  technique was  necessary to improve {\em
uv}-coverage and reduce bandwidth smearing. This technique  makes a
single image from multi-frequency data by gridding each spectral channel
in its correct place in the {\em uv} plane, instead of at a location
determined by the average over all channels. The {\em MIRIAD}
implementation of multi-frequency synthesis also solves for spectral
index, so that the image is not degraded by differing spectral indices
of sources in the field.

Before imaging, the data from each observing session were inspected and
the {\em MIRIAD} interactive tasks {\em TVFLAG}  and {\em BLFLAG} were used to flag bad
data resulting from interference, receiver problems or correlator
failures. The primary calibrator data were flagged before calibration
was applied.  The secondary calibrator and target data had bandpass and
polarisation calibration applied before inspection and flagging.

After flagging, the data were split into the separate observing bands
before imaging and cleaning, so that calibration and imaging could
account for frequency-dependent terms. When imaging, we explored a range
of robust weights,  which is a hybrid form of uniform and natural
weighting \citep{briggs95}.  Robust values of 1 or more resulted in a
degraded beam shape, which  translated into a strong circular artefact
in the final image. Decreasing  robust values gave tighter main lobes,
and hence higher resolution, better  clean models and more effective
self calibration, but at the cost of  sensitivity. A robustness value of
0 was chosen for the final imaging,  as it was found to be a good
compromise between sensitivity and resolution. This resulted in a
synthesized beam of 7.1 $\times$ 6.2 arcsec at 20 cm.

Both the \fbb\ and \fcc\ images contained sidelobes from two strong
off-field sources. The sidelobes in the  \fcc\ image were removed by
increasing the image size to image an area four times larger, and
cleaning  a small region, $\sim$2 arcmin square, around each source.
These clean components were removed from the visibility data before  the
final \fcc\ imaging. The \fbb\ off-field sources were removed by
offsetting the image region to include the two sources. The procedure
then followed that at \fcc\ to remove the source sidelobes.  

We experimented with a variety of approaches to deconvolution,
including, in the case of the \faa\ data, cleaning a region around the
bright central source only, and cleaning `boxes' around all bright
($>$ 1 mJy) sources. The number of clean iterations was chosen by
monitoring the  residuals of the clean model. 

The 20 and 11 cm data were self-calibrated to improve the  image
quality. The best self-calibration result  was from a clean model
derived over the whole image, with the number of iterations set as above
($\sim$ 30000  iterations). Two iterations of both phase and amplitude
self  calibration were sufficient to improve the  image quality, and in
particular remove phase error stripes.

After correction for the primary beam attenuation, image analysis was confined to a circular region of radius 20, 12, 5.5, and 3.5 arcmin at 20, 11, 6, and 3 cm respectively, at which radius the sensitivity falls to 39\% of that in the primary beam centre. The final rms noise at the image centre is shown in Table 2. A further indication of the noise statistics is given by measuring the deepest negative noise peak in the resulting image. Within the 6.5-arcmin circular region shown in Fig. 2, the most negative noise peaks are -47, -99, 
-76, and -84 \mj\ at 20, 11, 6 and 3 cm respectively.

\subsection{Source Extraction}
To identify sources in the image at each wavelength, we first divided each map by the noise map generated by SExtractor \citep{bertin04} to obtain a `signal to noise' 
map. The {\it MIRIAD} task {\it IMSAD} was then used to derive a preliminary 
list of source `islands' above a cut-off of $4\sigma$.

Each source `island' found by {\it IMSAD} was examined and refitted with an elliptical 
Gaussian to derive source flux densities and sizes. All sources and fit parameters were 
inspected to check for obvious failures and poor fits that need further analysis. 
The peak flux from this fit was then compared with a peak value derived from a parabolic fit to the source ({\it MIRIAD} task {\it MAXFIT}). If these two peak values differed by less than 20\% and the fitted position was inside the 0.9$S_{\rm peak}$ flux density contour, then these fluxes and positions were assigned a quality flag of 1. If they failed this test but inspection showed that the derived position and peak flux density of the peak were consistent with the data, then they were assigned a quality flag of 2. 

Cases which failed this second test fell into one of the following categories:

\begin {itemize}
\item{Sources which were better described by 
two Gaussians. In this case, the {\it IMSAD} islands were split into two Gaussian components, which are catalogued separately in Table 3.}

\item{Sources with a shape which could not be fitted by a Gaussian. No such sources are amongst those discussed in this paper.}

\item{Obviously spurious sources that correspond to imaging artefacts. These very rare cases were deleted.}

\end{itemize}

Finally, to reduce the incompleteness of our catalogue and minimise the probability of spurious detections, we included only sources with signal to noise greater than 5$\sigma$ in the final 
catalogue and in Table 3. 

\subsection {The limiting sensitivity of synthesis observations}
The radio images described here are amongst the deepest made with the ATCA. However,  future observations will probably be able to reach even deeper sensitivity levels, because of the planned introduction of a wide-bandwidth correlator (Compact Array Broadband Backend, or {\em CABB}) at the ATCA in 2006, and  through the potential future availability of even larger amounts of observing time for a few key science projects.

The sensitivity to compact sources achievable with a synthesis telescope is limited by four factors.

\begin{itemize}
\item {\bf{Thermal Sensitivity} 
\rm For many observations, particularly those with short integration
times, images are limited by the thermal noise temperature of the telescope,
which is set by the receivers, optics, and atmosphere. In the case of a
deep observation, this is still likely to be the limiting factor at high
frequencies, but  will become less important than other factors for deep
integrations at low frequencies. For multi-day observations with the
ATCA, this will continue to be the limiting factor at 6 and 3 cm,
because of the low confusion levels together with the faintness and low
surface brightness of sources at these frequencies. For example, the
ATCA can be expected to reach an rms noise level of 5 \mj\ after
observing at 6 cm for 16 *12 h (using two 128-MHz bands, and natural
weighting). This level depends primarily on the system gain and
temperature, and confusion does not play a significant role. This
sensitivity limit is often referred to as the 'theoretical limit'.}

\item {\bf{Confusion by adjacent sources} 
\rm Long integrations at low frequencies (20 and 13 cm) are likely to
reach the confusion limit at which each source of interest is confused
by an adjacent background source.  Assuming the log N/logS function
taken from \citet{jac04}, the confusion level, for a 6 arcsec beam at
20 cm, is 0.05 \mj/beam (i.e. on average, one background 0.05 \mj\ source will fall within each beam). Thus this does not currently present a limit to synthesis observations for any realistic amount of observing time.

However, a commonly stated rule-of-thumb is to require that there be no
more than one confusing source per 100 synthesised beams at the same
level as the target source, implying that most sources are separated by at least 10 synthesised beams. For the 6-km configuration of the ATCA,
using the same source statistics measured by Jackson (2004), and assuming a spectral
index of $-$0.7, this gives a practical sensitivity limit of 21, 6, 0.7,
and 0.03 \mj\ at 20, 13, 6, and 3 cm respectively. The 20 cm
observations described here already approach this limit. We note that
this rule-of-thumb is probably too conservative for deep-field surveys
such as those presented here, in which valuable astrophysical
information can still be gained even when, in some cases, there is some
confusion between adjacent sources.}

\item {\bf{Dynamic range limited by incomplete u-v coverage} 
\rm Even with perfect calibration, the dynamic range of an image is
limited by incomplete u-v coverage, and hence missing information, which
cannot be recovered even by sophisticated deconvolution algorithms. For
a carefully chosen field around a bright radio source, with no strong
confusing sources nearby, this is likely to be the factor that
ultimately limits the sensitivity. The highest dynamic range (defined as peak flux in the image divided by the rms in a source-free region of the image) ever achieved in
a synthesis image is  $10^6$, achieved using excellent {\em uv } coverage and redundant spacings on the Westerbork Synthesis
Radio Telescope \citep{bru05}. The highest dynamic range reached with the
VLA is $3.5 \times 10^5$ \citep{con93}, and a dynamic range of $10^5$ has been reached by
\citet{gel00} with the  ATCA. 

The observations presented here, which also have a dynamic range of $\sim 10^5$,  use a wide range of configurations, resulting in a uv coverage comparable with that of the VLA or Westerbork, and should therefore be capable of achieving a similar dynamic range. The images presented in this paper generally reach the thermal noise limit, and so are not yet capable of exploring much higher dynamic ranges. 

We note that the ATCA is in principle capable of almost complete {\em uv }
coverage,  by observing at every possible baseline, using a special set of 12 configurations \citep{man84}, which should in principle yield a very high dynamic range. 
However, the standard range of configurations currently offered to users does not include this special set. No deep ATCA observation has yet been demonstrably limited in dynamic range by incomplete {\em uv} coverage, but we note that future deep-field observations may require the use of such special configurations to reach the ultimate sensitivity.

In principle, even with a standard set of configurations, the {\em uv } coverage may be significantly improved by the use of the multi-frequency synthesis described above \citep{sault94}. However, although the   {\em MIRIAD} implementation of multi-frequency synthesis solves for spectral index, it still assumes a constant spectral index for each source over the observed frequency range. Data are not yet available to determine whether this will  limit the dynamic range in deep field observations.}

\item {\bf{Dynamic range limited by calibration errors} 
\rm The dynamic range will be limited by calibration errors. While
simple antenna-based complex gain errors can be corrected by the selfcal technique available in  {\em MIRIAD} and other packages, these implementations are unable to correct for gain errors that vary across the primary beam. Particularly troublesome are strong confusing sources close to the steep edge of the primary beam response (or its sidelobes), since small pointing errors may cause large amplitude fluctuations in these sources, preventing them from being adequately deconvolved from the image. Furthermore, the primary beam response is assumed in current analysis packages to be circularly symmetric. 

The magnitude of this effect is uncertain, but an indication of its severity may be estimated as follows. Assuming that the critical area of the primary beam at 20 cm in which pointing errors will be significant occupies an annulus of radius 15 arcmin and width 5 arcmin, then we expect one 40 mJy source to fall by chance within that annulus. A typical dynamic range of 500 per 12-hour synthesis for sources in this region will then limit the sensitivity to 40 \mj\  for each 12-hour image. When n observations are combined, then if the pointing errors are purely random, this will be reduced by sqrt(n), giving a limiting rms sensitivity of 10 \mj\ for the set of observations described in this paper. Corresponding limits for 11 and 6 cm are 4 and 1 \mj\ respectively. This limiting sensitivity at 20 cm is similar to the sensitivity obtained in the observations described here, suggesting that more sophisticated calibration algorithms may be necessary to enable significantly deeper imaging. 
}
\end{itemize}

If future observations wish to probe more deeply in the presence of such
strong confusing sources, then it will be necessary to develop calibration packages that account for varying gain errors
across the primary beam. We note that the aips++ package contains an algorithm that can in
principle handle these errors.

\section {Radio Sources in the Inner HDF-S Region}

The final 20 cm image is shown in Figure 1. Overlaid on this image
are the primary beam sizes and locations for the ATCA observations at the four
frequencies, and the locations of the HST WFPC, STIS, and NICMOS fields.
Of particular note are the bright confusing source
ATHDFS\_J223355.6$-$604315 at 6  arcmin south-east of image centre, the
clearly identifiable multiple radio source in the north-east corner, and
a radio galaxy with a jet-lobe structure about 7 arcmin south of the
image centre. 

Catalogues covering the complete region imaged by the AT at the two
longest frequencies will be presented in Papers 2-4. All the radio data
from the survey are available on \\
http://www.atnf.csiro.au/research/deep/hdfs/, and in the NASA/IPAC
Extragalactic Database (NED) on
http://nedwww.ipac.caltech.edu/.

In this remainder of this paper we focus on a circular region of radius 6.5
arcmin, centred on the WFPC field at (RA =
22h 32m 56.22s and Dec = $-$60$^\circ$ 33' 02.7'' (J2000) ). The size of this region has been
chosen to include  the HST WFPC2, NICMOS and STIS fields, and also cover
all of the area inside the \fcc\ and \fdd\ observations.  Thus all the
6 and 3 cm data from this survey are presented in this paper. The
\faa\ image of this region is shown in Figure 2.

In Table 3 we show the combined source catalogue for all sources
detected at a level of 5 times the local rms noise at one or more
frequencies within this region. Sources were matched across the 4
frequencies by visual inspection and the result is a catalogue of 87
radio sources.  For  sources detected at more than one wavelength, the
position is derived from the image with the smallest errors in the
source position, as noted in column 15 of Table 3.

The columns of Table 3 are as follows.

{\noindent{\em Column 1} --- Reference number. These are used for
brevity only within this paper, and, to avoid ambiguity, the full source
name should be used by any papers referring to the catalogue.}

{\noindent{\em Column 2} --- Source name.}

{\noindent{\em Columns 3 \& 4} --- Right Ascension (J2000), 
1$\sigma$ uncertainty in arcsec. Note that these uncertainties refer to
the formal positional uncertainties derived from the fitting process. To
these should be added, in quadrature, a systematic uncertainty of about
0.2 arcsec, representing the uncertainty in the position of the AT phase
calibrator sources. }

{\noindent{\em Columns 5 \& 6} --- Declination (J2000),  
1$\sigma$ uncertainty in arcsec.}

{\noindent{\em Columns (7--10)} --- Flux densities at 
20, 11, 6, and 3 cm in \mj. Where the column is blank, then
the source is either undetected at that wavelength, i.e.  below $5\sigma_{local,\nu}$,  or lies  outside the catalogued area. The peak flux is given for unresolved sources and the integrated flux is given for resolved sources.}

{\noindent{\em Column (11)} --- Deconvolved major axis (FWHM), 
$\Theta_{maj}$ in arcsec, at the wavelength indicated in column 15. }

{\noindent{\em Column (12)} --- Deconvolved minor axis (FWHM),  
$\Theta_{min}$, in arcsec, at the wavelength indicated in column 15.}

{\noindent{\em Column (13)} --- Deconvolved position angle, PA, in
degrees
east of north, at the wavelength indicated in column 15. }

{\noindent{\em Column (14)} --- Local signal to noise ratio.}

{\noindent{\em Column (15)} --- Band flag - indicates the wavelength
at which the source position and Gaussian fit 
parameters are measured: L= \faa, S= \fbb, 
C= \fcc, and X= \fdd.}

{\noindent{\em Column (16)} --- Quality flag, as described in Section 3.2}

\section{Properties of Selected Radio Sources}

In this section, we discuss in detail a subset of 19 representative
sources chosen from the list given in Table 3. These sources
have not been selected in any statistically well-defined way, and so
conclusions drawn from this subset may not be true for the population
of sources as a whole. The full sample will be discussed in detail in
Paper 3. 

It is particularly important to determine the origin of the radio luminosity, which may be generated either by star formation or AGN activity. A potential discriminator between these two mechanisms is morphology. At higher flux densities, AGNs frequently have a classic double-lobed structure, whilst starburst galaxies are typically unresolved or amorphous. However, the extent to which these morphologies are present in \mj\ sources is not yet well-established, and the resolution of the observations presented here is in many cases insufficient to distinguish between these. 

Another potential discriminant is radio spectral index. Starburst galaxies typically have a spectral index of about $-$0.7, whilst AGNs typically have a spectral index ranging from about 0 to about $-$1.4. Thus, while little is learnt from a spectral index of about $-$0.7, a spectral index which differs significantly from this, in either direction, implies AGN activity. However, the spectral indices presented here have a typical uncertainty, due to image noise, of $\sim$ 0.1 -- 0.2, so only extreme spectral indices are useful indicators. There is a further uncertainty caused by the mismatch of the synthesised beam at different frequencies. \citet{P3} reduce this uncertainty by convolving the higher-frequency image with the lower-frequency beam, and we use their spectral indices here, which therefore differ slightly from those which would be obtained from the figures in Table 3.

One particular class of radio source, the ``gigahertz peaked spectrum" or GPS galaxy, has a spectrum which rises to a maximum at centimetre wavelengths,  and is believed to represent an early stage of AGN activity \citep{sne99}.  This characteristic spectrum is seen in two of the objects discussed here, and is an unambiguous indicator of AGN activity.

A potential cause of a flat radio spectrum might also be that the shorter wavelength emission contains a component of free-free emission from HII regions in a  starburst galaxy. However, the spectral indices discussed here are based on the 20 and 11 cm radio emission, at which wavelength it would be unusual for free-free emission to be significant \citep{con92}.

A further complication for any interpretation of these sources is that a starburst galaxy may  contain an obscured AGN in its nucleus. Local examples of this phenomenon are  NGC6240 \citep{gal04} and Mrk231 \citep{yun01} in which a starburst galaxy harbours a hidden
low-luminosity AGN, with comparable contributions to the luminosity from
the starburst and AGN components.  

A list of the sources and their optical, infrared, and derived
properties is given in Table 4. In Figure 3 we show the \faa\  radio
contours overlaid on CTIO images \citep{pal00}. The columns of Table 4 are as follows.

{\noindent{\em Column 1} --- Reference number, as in Table 3.}

{\noindent{\em Column 2} --- Source name.}

{\noindent{\em Column 3} --- Flux density at \faa.}

{\noindent{\em Column 4} --- Measured redshift. Redshifts with four
decimal digits are spectroscopic redshifts, while those with two decimal
digits are photometric redshifts. References for the redshifts are given
in the discussion of individual sources. For reasons discussed in \citet{P3},  in cases where both
\citet{FS} and \citet{tep01} have measured photometric redshifts, but no
spectroscopic redshifts are available, we adopt the \citet{FS} redshift.}

{\noindent{\em Columns 5--9} --- AB magnitudes measured at V, R, I, J,
and K bands. Except where otherwise stated, measured values of V,R,I are
taken from the CTIO catalogue \citep{pal00}.  Other references are given
in the discussion of individual sources.}

{\noindent{\em Column (10)} --- I-K colour.}

{\noindent{\em Column (11)} --- Radio luminosity}

{\noindent{\em Column (12)} --- Star formation rate implied by the radio
luminosity, if all radio emission is assumed to be generated by star formation activity, using the scaling given by \citet{con92}  with a
Saltpeter IMF (Q=5.5). For comparison, this algorithm assigns Arp220  a
star formation rate of 300 \myr .}

{\noindent{\em Column (13)} --- Logarithm of the radio to optical
luminosity ratio, calculated as $0.4 \times (I -S_{20})$, where I is the
I-band magnitude shown in Table 2, and $S_{20}$,  is the radio magnitude
derived from the \faa\ flux density using the AB Radio magnitude scale
defined by Ivezic et al. (2002). In a few cases, indicated by the notes
to Table 2 and in the discussion of individual sources, I has been
estimated from R or J magnitudes. For comparison, Arp220 has log
($S_{20}/I) = 1.31$  }

{\noindent{\em Column (14)} --- Classification of the radio source. The
arguments for each classification are given in the discussion of
individual sources. SB indicates starburst activity, Sy indicates a
Seyfert galaxy, and GPS indicates a gigahertz-peaked-spectrum AGN.}

The rest of this section discusses each of the sources in Table 4. In
each case, the discussion is headed by the short reference number, the
full source name, and the source name given by Norris et al. (1999,
2001), if appropriate. We use a number of abbreviations throughout this
section: FS refers to \citet{FS}, and SMO refers to \citet{smo}. We
refer frequently to the CTIO images \citep{pal00}, the WFPC images
\citep{wil00, cas00}, the WFPC FF (flanking field) observations
\citep{luc03}, the NICMOS images \citep{yah00}, and the EIS (ESO
Infrared survey; \citet{non99}). Except where stated otherwise, all radio spectral indices are taken from \citet{P3}, and refer to the spectral index between 20 and 11 cm.

\subsection*{ 3. ATHDFS\_J223245.6-603857 (source b) }
This bright, marginally resolved, 843  \mj\ source has a radio spectral
index of ($-0.69 \pm 0.06$), which reveals little about the origin of its radio luminosity. The source is located about one arcsec east of a faint, possibly
disturbed or interacting, galaxy seen in WFPC FF and, faintly, in CTIO
images.  Here we assume they are coincident, because astrometric
accuracy on the flanking fields is relatively poor. FS fit an irregular
template and obtain a photometric redshift of 0.75. If the resulting
high radio luminosity is attributed entirely to star formation activity,
this implies a star formation rate of  515 \myr. The radio-optical luminosity ratio
is also very high, implying a high degree of extinction. One explanation is that it is a very obscured starburst galaxy, about twice as active as Arp220, similar to those suggested by \citet{bar00} to account for the strong SCUBA sources in HDF-N, and an alternative is that the source contains a hidden AGN. Given the irregular spectral
template, and the disturbed morphology, it is likely that this is a very
obscured starburst galaxy, although we cannot rule out the presence of an obscured AGN.

\subsection*{ 7. ATHDFS\_J223254.5-603748 }
This faint 92  \mj\ source is detected only in our \faa\ observations.
It is coincident with a bright disturbed barred spiral galaxy seen in
CTIO and WFPC FF images. \citet{P4} measure a spectroscopic redshift of
0.1798, and see a Seyfert-type spectra. Thus, this source appears to be
a relatively normal Seyfert galaxy. We note that SMO have measured a
spectroscopic redshift of 0.2668, but the reason for this discrepancy is
unclear. Here we adopt the \citet{P4} redshift of 0.1798. 


\subsection*{ 12. ATHDFS\_J223316.5-603627 (source f) }
This bright, unresolved, 649  \mj\ source has a radio spectral index of
($-0.67 \pm 0.06$). It is coincident with a galaxy in the CTIO image that appears
extended, and is possibly an edge-on spiral.  \citet{tep01} measure a
photometric redshift of 0.60, while FS measure a photometric redshift of
0.29, using a starburst template. Here we adopt the FS redshift. The
radio luminosity then implies a star formation rate of 95 \myr. This,
together with the colours of this galaxy, indicates vigorous, but not
unusual, starburst activity, in a relatively normal galaxy.

\subsection*{ 19. ATHDFS\_J223338.8-603523 ( source m) }
The radio spectrum of this 185 \mj\ unresolved source is relatively flat
($-0.35 \pm 0.21$) suggesting an AGN rather than a starburst galaxy. The source is coincident with a bright elliptical galaxy in the WFPC FF image,  which
is also seen as a J=18.3 galaxy in the EIS survey, with J-K=0.54.
\citet{P4} have measured a spectroscopic redshift of 0.2250, and we note
that \citet{tep01} have measured a photometric redshift of  z=0.16. The
properties of this source are consistent with those of an elliptical galaxy hosting an AGN.

\subsection*{ 26. ATHDFS\_J223243.3-603442 }
This faint radio source is visible in our \faa\ and \fcc\ observations,
but not in the \fbb\ observations, which give an upper limit \fluxb\ $<$
30\mj. It has a radio spectral index derived directly from the 6 and 20 cm
observations of $-$0.23, suggesting that it is an AGN, although this conclusion is weakened by the faintness of the source, its non-detection at 11cm, and the different beam sizes at 6 and 20 cm. The source is coincident
with an extended and apparently disturbed system, possibly a merger,
visible in CTIO and WFPC images.  \citet{tep01} measured a photometric
redshift of  0.57, FS measured a photometric redshift of 0.46, and SMO
measure a spectroscopic redshift of 0.4233, which we adopt here. \citet{man02} have observed it at 7 and 15 $\mu$m with ISO, and show that
its spectral energy distribution (SED) resembles a cirrus-dominated
galaxy like M51 rather than a starburst. Subject to the uncertainty discussed above, the radio spectrum suggests that this otherwise normal spiral galaxy may harbour a weak radio AGN in its nucleus.

\subsection*{ 29. ATHDFS\_J223245.5-603419 (source a) }
This bright radio source has a spectral index of ($-0.43 \pm 0.21$), which is marginally flatter than that expected of a starburst galaxy. It is identified 
with the western-most member of a line of three galaxies clearly visible in the
CTIO, AAT and EIS images, but unfortunately just two arcsec outside the
deep WFPC field. Its radio image is slightly extended in the direction
of the line of galaxies, suggesting that these may contribute weakly to
the radio emission. \citet{fra03} suggest that most of the
infrared emission is from the central object of the triplet, and yet the
position given in their table coincides with the western object, which
is unambiguously the origin of the strong radio emission. We list their
J and K magnitude in Table 2. \citet{gla98} have measured a
spectroscopic redshift of 0.4606, while \citet{van02} measure a
redshift of 0.4594, and SMO measure z=0.4608. We adopt the mean of
these, which is 0.4603.  \citet{tep01}  and FS have measured
photometric redshifts of 0.57 and 0.46 respectively, and FS fit an sbc
template to this galaxy. Based on its infrared properties, \citet{fra03} derive a mass of $10^{11} M_{\sun} $ and a star formation rate of  50 \myr, which is slightly lower than the rate of 81 \myr that we derive from the radio observations. \citet{man02} have observed
it at 7 and 15 $\mu$m with ISO, and show that its spectral energy
distribution (SED) resembles a starburst galaxy like Arp220. The data presented here are consistent with this interpretation.

\subsection*{ 31. ATHDFS\_J223327.6-603414 (source h) }
This spectrum of this bright unresolved radio source peaks at \fbb, and
we therefore classify it as a GPS galaxy,
which is believed to represent an early stage of AGN activity
\citep{sne99}. There is no optical counterpart, but it lies about two
arcsec from a small galaxy (at 22 33 27.40, -60 34 12.37) which is
visible within the STIS flanking fields observations. The absence of any
galaxy coincident with the radio source in the HDF flanking field observations gives limiting magnitudes shown in Table 4, leading to an
extreme value of the radio-optical luminosity ratio, indicating that the
GPS source is strongly obscured by its host galaxy.

\subsection*{ 33. ATHDFS\_J223243.4-603351 }
This unresolved radio source has a flux density of 98  \mj\ at \faa\ ,
rising to 114  \mj\ at \fdd\ .  This inverted spectrum suggests that it
is an AGN. It is coincident with a point-like optical source on the CTIO
image, and \citet{P4} have measured a spectroscopic redshift of 1.566,
and note that it has the broad lines of a quasar. \citet{pal00}
(who refer to it as QSO B) and \citet{fra03} also identify
this source as a quasar at z=1.56. \citet{man02} have observed it
at 7 and 15 $\mu$m with ISO, and suggest that it is a broad-line AGN,
but measure a redshift of 0.0918 based on a single broad line. As this
differs from the redshifts measured by \citet{P4}, Palunas et al., and
Francheschini et al., we suggest that the Mann et al. redshift was based

on a misidentified line, and adopt a redshift of 1.566 for this quasar. 

\subsection*{ 34. ATHDFS\_J223306.0-603350 (source d) }
This bright radio source has a spectral index of ($-0.62 \pm 0.12$), and is
identified with the core of a relatively bright face-on barred spiral
galaxy (R=17.2), which appears to be interacting with a small galaxy a
few arcsec to its north. The radio source is significantly extended to
the south at \fbb, possibly indicating some structure within the galaxy.
\citet{gla98} have measured a redshift of 0.1733, and show that it has
the characteristic emission lines of a star-forming galaxy. It is the
brightest 15-$\mu$m source seen in the ISO observations by \citet{man02}
 who show that its SED is that of a normal spiral galaxy rather
than a starburst. However, its optical and radio luminosities suggest
that it is a luminous starburst galaxy, and Francheschini et al. show
that its infrared emission shows an excess over that expected from a
normal spiral. We derive a star formation rate of 27 \myr, and thus
classify it as a starburst rather than a normal spiral.

\subsection*{ 35. ATHDFS\_J223258.5-603346 (source c) }
This unusual source is the strongest radio source in the WFPC field, and
yet is optically a very faint (V=27.05) red source which is  invisible
in CTIO images and barely visible in WFPC images. It has been discussed
extensively by \cite{nor99} and \cite{nor01}, where it is labelled as
"source c". A number of authors have measured photometric redshifts for
this galaxy: FS (1.69), \cite{fon04} (1.7), \citet{lan02} (1.69),
\citet{rod01} (1.69), and \citet{rud01} (1.34). We note the high
degree of consistency between these measurements, and adopt a redshift
of 1.69 for this object. A 12-hour spectroscopic observation on the VLT
failed to detect any emission lines \citep{van02}.  

This source has an unusually high radio-optical ratio (the highest in
this subset), several hundred times greater than Arp220.  \citet{van01} measure  I-K$_{AB}$ = 3.45, classifying it as an extremely red
object (ERO) according to the division specified by 
\citet{poz00}. EROs \citep{mcc04} are thought to be a mix of passively
evolving red galaxies at $1 < z < 2$ and heavily obscured star-forming
galaxies, also at z $>$ 1. However, as EROs include both starburst and
early-type galaxies, this classification does not yield any further
information on this object. Based on the infrared colours, \citet{van01} also classify it as an elliptical galaxy. It appears to be marginally
extended in both  the 3 cm image and the WFPC image. 

The radio spectral index of this source is ($-0.65 \pm 0.05$). It is difficult to determine whether this source is a extremely dusty
starburst or a dust-enshrouded AGN. Its radio spectral index is
consistent with either, and SCUBA observations of similar galaxies in
the HDF-N \citep{bar00} suggest that their radio emission is produced by
star formation. At a redshift of 1.69, this would imply this source is
about five times more luminous in the radio than Arp220.  On the other
hand, unpublished SIMBA \citep{nym01} observations by Wiklind, Bergstrom, Huynh,
Norris, and Jackson failed to detect any 1.3 mm continuum emission greater than 7.5 mJy, whereas if this source is powered by starburst
activity, the associated dust would be expected to have a flux density
of $>$ 15 mJy. We therefore consider it likely that this source is a
dust-enshrouded AGN.

\subsection*{ 37. ATHDFS\_J223247.6-603337 }
This faint radio source is seen only in our 20 cm observations, and is
coincident with a large late-type spiral galaxy in the CTIO image.
\citet{gla98} measure a spectroscopic redshift of 0.5803, and SMO
measure a spectroscopic redshift of  0.5807. Photometric redshifts of
0.67 and 0.56 have been measured by \citet{tep01}  and FS respectively.
\citet{man02} note that its SED is that of a normal galaxy like
M51. \citet{rig02} and \citet{fra03} show
that, based on spectroscopy and infrared photometry respectively, it is
an unusually massive galaxy, with a dynamical mass of $4.5-10 \times
10^{11} M_{\sun} $ and a star formation rate of  45 \myr. This is in
good agreement with the star formation rate of 32 \myr\ that we derive
from its radio luminosity.

\subsection*{ 39. ATHDFS\_J223337.5-603329 (source l) }
This source represents the bright, unresolved, radio emission from the
well-studied STIS quasar \citep{out99} at z=2.238, whose
location was partly responsible for the location of the HDF-S, as
discussed in Section 2.1 above. This source has been extensively
discussed in the literature (e.g. \citet{sea98, pal00}), and so we
restrict the discussion here to noting that the associated radio source
is typical of a powerful radio-loud quasar, with a spectral index of ($-0.69 \pm 0.05$).

\subsection*{ 43. ATHDFS\_J223327.9-603304 (source i) }
This bright double radio source appears in Table 3 with two entries
(43. ATHDFS\_J223327.9-603304 and 44. ATHDFS\_J223329.2-603302  ) corresponding to the two lobes of the source,
indicating that it is a radio galaxy or quasar. There is no optical counterpart visible in the CTIO
images, but a faint counterpart is visible at J and K in the  EIS
images. The limit on I from the CTIO catalogue gives a radio-infrared
luminosity ratio of 5.20, indicating that the source is highly obscured. 

\subsection*{ 48. ATHDFS\_J223308.6-603251(source e) }
This strong extended radio source is coincident with a R=21.77 CTIO
spheroidal galaxy visible in the CTIO, AAO, and WFPC FF1 images. Its
radio spectral index is ($-0.86 \pm 0.03$), suggesting a radio galaxy rather than a
starburst. The radio source appears to be extended in a NW-SE direction
at 20, 11, and 6 cm by  about 5 arcsec, corresponding to 10 kpc at
z=0.5. At \faa\ our spatial resolution is insufficient to tell whether this extension may be partly due to contamination by the emission from a nearby galaxy, but at \fdd\ it appears as a core-dominated source with weak jets
extending to the NW and SE. Both \citet{tep01} and FS measure
photometric redshifts of  0.64. All these data are consistent with a
classical radio-loud AGN of moderate radio power.

\subsection*{ 49. ATHDFS\_J223323.2-603249 (source g) }
This 457 \mj\ source is notable in having a GPS
spectrum, suggesting that it is powered by AGN activity. It is marginally resolved at \fbb\, and unresolved at other wavelengths. No optical
counterpart is visible in the CTIO image of this field, but it is weakly
detected at J and K by EIS. For the calculation of the
radio-optical ratio, the I magnitude has been estimated as I=J+0.12,
where 0.12 is the mean value of I-J for those sources in this subset
for which a measured value of I-J is available. This source has one of
the largest radio-optical ratios in our subset.

\subsection*{ 55. ATHDFS\_J223331.6-603222 (source k) }

This slightly  resolved 295 \mj\ source has a spectral index of ($-0.68 \pm 0.15$),
which gives no clue to the origin of its emission. It  is coincident
with the central galaxy of a group of three galaxies, possibly merging,
seen in the CTIO, AAO, WFPC-FF, and EIS images, with J-K= 0.8. FS
measure a photometric redshift of 0.42, and \cite{P4} measure a
spectroscopic redshift of 0.4652, with the spectrum characteristic of a
star-forming galaxy. Its extension at 20 and 13 cm is consistent with
some radio emission from the other two galaxies in the group. Its radio
luminosity implies a star formation rate of 123 \myr, and its
radio-optical ratio is about ten times that of Arp220. These
observations are consistent with a very obscured starburst galaxy.

\subsection*{ 67. ATHDFS\_J223236.5-603000 }
This is the strongest (1.5 mJy) radio source in the subset discussed in
this paper. It has a spectral index of ($-0.86 \pm 0.05$), and no optical
counterpart. The EIS upper limits give it an extreme radio-optical ratio
of 6.02. It is marginally resolved at 20 cm, and also shows  a small
amount of emission to the south in the 13 cm image, which is coincident
with another galaxy, and may be unrelated. None of the observed
characteristics gives any clue as to whether this radio emission is
powered by AGN or star formation activity, but the high radio-optical
luminosity ratio indicates a high obscuration by dust, whatever the
origin of the radio emission.

\subsection*{ 69. ATHDFS\_J223316.8-602934 }

This extended strong (1.0 mJy) source has a flat spectrum (spectral
index ($-0.16 \pm 0.06$)), indicating that it is an AGN. It is coincident with a
faint galaxy in the CTIO image, in a dense cluster. Photometric
redshifts of 0.12 and 0.89 have been obtained by FS and \citet{tep01}
respectively. Here we adopt the FS redshift. Some of the extended radio
emission may be attributable to the other galaxies in the cluster.

\subsection*{ 70. ATHDFS\_J223329.1-602933 }
This extended 261 \mj\ source appears in our source catalogue only as a
\faa\ detection, although there is a marginal detection in the \fbb\
observations with a peak flux density of 86  \mj, giving a spectral
index based on this uncorrected flux of $-$1.9, indicating an AGN.  We note that this would be classified
as an ''ultra-steep spectrum'' by \citet{deb00}, and consequently may be
a high redshift candidate. A faint extended source, or cluster of
sources, is visible in the CTIO image, but no redshift is available.
Although V and I magnitudes are given in the  CTIO catalogue, they have
very high uncertainties and so have been set to zero in Table 4. For the
calculation of the radio-optical ratio,  an I magnitude has been
estimated as I=R$-$0.83, where 0.83 is the mean value of R-I for those
sources in this subset for which a measured value of R-I is
available. The extended radio emission to the north and east may
represent emission from the other galaxies in the cluster.

\section{Discussion}

Of the 19 sources in our subset, we classify eleven as AGN or
quasars, six as starbursts or star-forming galaxies, and two as
composite or unknown. However, the selection criteria for this
subset are not well-defined, and the subset may be biased towards
brighter objects, and so these fractions are unlikely to be
representative of the population as a whole. For example, \citet{pra01}
find that 40\% of sources $<$ 1 mJy are starburst galaxies. A careful
analysis of the distribution of the entire sample of sources from this
survey will be given by \citet{P4}.

About half of the objects in our subset have high radio-optical
luminosity ratios which appear to be quite unlike those seen in the
local Universe. However, a detailed comparison requires a significant
correction (the k-correction) to be made to the flux densities of these
objects, both at infrared and radio wavelengths, to compensate for the
fact that the observed wavelength differs significantly from the emitted
wavelength.  This correction depends sensitively on the intrinsic
emission spectrum of the object, and the redshift. Even in those cases
where redshifts are known, the intrinsic spectrum of the galaxy is
unknown, and so attempts to produce plots of  radio/optical luminosities
depend sensitively upon assuming a template derived from the local galaxy population,
which may be intrinsically different from those at high redshifts.  Small
differences in these spectra result in significantly different conclusions,
and so the resulting plots may be misleading. Thus we are unable to make a
detailed comparison of radio-optical ratios with those seen in the local
Universe until these objects are much better characterised.

We also note that our lack of knowledge of the intrinsic spectra of
these objects is likely to make photometric redshifts unreliable. In the
detailed discussion of Section 5, we have noted several instances where a
photometric redshift differs significantly from a spectroscopic
redshift. However, the reliability of photometric
redshifts for the heavily obscured objects will increase as high-quality
mid-infrared observations become available from the Spitzer Space
Telescope, together with the detailed models and templates based on
those observations.

A significant fraction of the objects in this subset appear to be
heavily reddened, and fall into the category of ``Extremely Red Objects"
(EROs). A similar result is found in other Deep Fields observations such
as those in the Hubble Deep Field North (e.g. \citet{gar00},
\citet{ric00} ). For example, \citet{wad99} detected a faint radio
galaxy in the HDF-N (VLA\_J123642$+$621331) at a redshift of 4.42, which
in several respects may be similar to some of the objects described
here. Numerous observations of the EROs at a variety of wavelengths have
established that this class includes both early-type evolved galaxies at
z $>$ 1, heavily obscured high -- z active galaxies and starburst
galaxies, and in some cases a combination of these (e.g. \citet{afo01},
\citet{bru02}, \citet{ber04}, \citet{geo04} ).

\section {Conclusion}
In this first paper of the series, we have presented the observations
and described some of the radio objects detected in the Hubble Deep
Field South. Subsequent papers will present the entire catalogue and
discuss identifications, source properties, and the statistical properties of the sample. 

The small subset of galaxies described here show that:
\begin{itemize}

\item The galaxies include a mixture of starburst-dominated and
AGN-dominated, with some evidence that the luminosity of some galaxies
may be generated by a combination of these phenomena.

\item Several of the galaxies have unusually large radio-optical
luminosity ratios, and  appear to be heavily obscured by dust.
\end{itemize}

Whilst the subset discussed in detail here has not been chosen according
to any statistically well-defined procedure, it is already clear from
this small sample that some of these galaxies are quite unlike those
seen in the local Universe, perhaps representing an earlier evolutionary
stage. This has already been noted by other authors conducting deep
radio surveys, but the small numbers of such objects makes it difficult
to  characterise these objects. It is also likely that many of these
objects will be more clearly characterised by mid-infrared results from
the Spitzer Space Telescope.

\section {Acknowledgements}
We thank Tommy Wiklind and Stefan Bergstrom for their collaboration in the unpublished SIMBA
Observations, and we thank Jose Afonso for valuable comments on an earlier draft of this paper. The Australia Telescope Compact Array is part of the Australia Telescope which is funded by the Commonwealth of Australia for operation as a National Facility managed by CSIRO.

\clearpage

\setcounter{table}{0}

\begin{table}
\caption{The Fields Searched and the resulting peak flux and rms in each
field
}

\begin{tabular}{ rcccrr }
 Field & HST Candidate    & 
\multicolumn{2}{c}{Pointing centre (J2000)} & Peak Flux & rms \\
Name & Fields      & 
RA         & Dec       & (mJy) & \ (mJy) \\

\hline
\\

1A/B &  1A \& 1B  &  22:50:13.92 &  -60:19:26.2  &  53 &  0.097 \\
3A/QSO &  3A \& QSO  (see note)  &  22:33:18.63 &  -60:36:52.0  &  155 &
0.116 \\
4A &  4A  &  22:13:33.47 &  $-$61:02:03.0  &  69 &  0.090 \\
15A/B &  15A \& 15B  &  23:30:44.28 &  $-$60:28:17.0  &  42 &  0.087 \\
12A/B &  12A \& 12B  &  00:37:44.31 &  $-$62:18:37.8  &  179 &  0.193 \\
11A/B &  11A \& 11B  &  02:51:22.97 &  $-$60:34:36.0  &  59 &  0.102 \\
16A &  16A  &  02:49:28.97 &  $-$63:30:46.5  &  21 &  0.087 \\

14A &  14A  &  04:12:58.98 &  $-$59:02:15.8  &  20 &  0.082 \\

\hline
\end{tabular}
\end{table}

\begin{deluxetable}{rccccc}
\tabletypesize{\small}
\tablewidth{0pt}
\tablecaption {HDF-S Pointing Centres and Total Observing Time}

\tablehead{
\colhead{Central}   & 
\colhead{Primary beam}    & 
\multicolumn{2}{c}{Pointing centre (J2000)} & 
\colhead{Total Hours} & 
\colhead{Central} 
\\
\colhead{frequency} & 
\colhead{FWHP/arcmin}      & 
\colhead{RA}         & 
\colhead{Dec }      & 
\colhead{per band} & 
\colhead{rms \ $\mu$Jy}
}
\startdata
 1.4 GHz  & 33 &  22:33:25.96 & $-$60:38:09.0 & 190 & 16.1 \\
 2.5 GHz  & 22 &  22:33:25.96 & $-$60:38:09.0 & 190 & 11.4 \\
 5.1 GHz  & 10 &  22:32:56.22 & $-$60:33:02.7 & 208 & 10.9 \\
 8.7 GHz  &  5 &  22:32:56.22 & $-$60:33:02.7 & 200 & 11.8 \\
\enddata

\tablenotetext{a}{Hours observed are given per 128-MHz band. These
numbers should be halved to give the observing time when observing with
two 128-MHz bands.}

\label{tab2}
\end{deluxetable}

\clearpage

\begin{deluxetable}{lllllllrrrrrrrrc}

\tabletypesize{\scriptsize}
\rotate
\tablewidth{0pt}
\setlength{\tabcolsep}{0.04in}

\tablecaption{Catalogue of ATHDFS sources within 6.5 arcmin of J2000 
22$^{h}$ 32$^{m}$ 56.22'', $-$60$^{\circ}$ 33' 02.7''. The columns are described in Section 4.}
\label{atcat}
\tablehead{
\colhead{Ref} & 
\colhead{Name} & 
\colhead{RA (J2000)} &
\colhead{$\sigma_{RA}$} & 
\colhead{Dec (J2000)} & 
\colhead{$\sigma_{Dec}$}& 
\colhead{\fluxa} &  
\colhead{\fluxb} & 
\colhead{\fluxc} &
\colhead{\fluxd} & 
\colhead{$\theta_{\rm maj}$} & 
\colhead{$\theta_{\rm min}$} & 
\colhead{PA} & 
\colhead{SN} & 
\colhead{band} & 
\colhead{qual}
}
\startdata
1 & ATHDFS\_J223253.7$-$603921 &  22:32:53.70  & 0.69 &  $-$60:39:21.2  & 0.62 & 0.052 & - & - & - & 0 & 0 & 0 & 5.1 & L &  2  \\ 
2 & ATHDFS\_J223258.7$-$603903 &  22:32:58.79  & 0.36 &  $-$60:39:03.9  & 0.47 & 0.058 & - & - & - & 0 & 0 & 0 & 5.7 & L &  2  \\ 
3 & ATHDFS\_J223245.6$-$603857 &  22:32:45.65  & 0.03 &  $-$60:38:57.3  & 0.04 & 0.843 & 0.536 & - & - & 2.27 & 0.73 & 30.5 & 75.9 & L & 1  \\ 
4 & ATHDFS\_J223307.1$-$603846 &  22:33:07.13  & 0.30 &  $-$60:38:46.9  & 0.46 & 0.137 & 0.074 & - & - & 4.9 & 2.15 & 17 & 9 & L &  2  \\ 
5 & ATHDFS\_J223304.6$-$603835 &  22:33:04.66  & 0.52 &  $-$60:38:35.8  & 0.52 & 0.067 & - & - & - & 0 & 0 & 0 & 5.6 & L &  2  \\ 
6 & ATHDFS\_J223248.2$-$603805 &  22:32:48.29  & 0.33 &  $-$60:38:05.6  & 0.44 & 0.076 & - & - & - & 0 & 0 & 0 & 7.4 & L &  2  \\ 
7 & ATHDFS\_J223254.5$-$603748 &  22:32:54.53  & 0.38 &  $-$60:37:48.0  & 0.48 & 0.092 & - & - & - & 3.85 & 2.43 & 42.4 & 7.7 & L &  2  \\ 
8 & ATHDFS\_J223232.8$-$603737 &  22:32:32.85  & 0.06 &  $-$60:37:37.1  & 0.08 & 0.645 & 0.489 & 0.216 & - & 3.63 & 2 & 19.6 & 45.8 & L & 1  \\ 
9 & ATHDFS\_J223225.6$-$603717 &  22:32:25.68  & 0.36 &  $-$60:37:17.4  & 0.84 & 0.057 & - & - & - & 0 & 0 & 0 & 5.3 & L &  2  \\ 
10 & ATHDFS\_J223326.0$-$603716 &  22:33:26.01  & 0.49 &  $-$60:37:16.7  & 0.95 & 0.061 & - & - & - & 0 & 0 & 0 & 5 & L &  2  \\ 
11 & ATHDFS\_J223236.6$-$603657 &  22:32:36.60 & 0.18 &  $-$60:36:57.7 & 0.44 & - & 0.095 & - & - & 0 & 0 & 0 & 7.5 & S &  2 \\ 
12 & ATHDFS\_J223316.5$-$603627 &  22:33:16.55  & 0.05 &  $-$60:36:27.5  & 0.05 & 0.649 & 0.451 & 0.204 & - & 0 & 0 & 0 & 56.2 & L & 1  \\ 
13 & ATHDFS\_J223232.5$-$603553 &  22:32:32.56  & 0.12 &  $-$60:35:53.1  & 0.18 & 0.466 & 0.236 & 0.141 & - & 6.82 & 4.54 &   $-$8.9  & 27.2 & L & 1  \\ 
14 & ATHDFS\_J223229.8$-$603544 &  22:32:29.89  & 0.35 &  $-$60:35:44.9  & 0.45 & 0.077 & - & - & - & 0 & 0 & 0 & 7.9 & L &  2  \\ 
15 & ATHDFS\_J223232.4$-$603542 &  22:32:32.40  & 0.12 &  $-$60:35:42.5  & 0.16 & 0.529 & 0.334 & 0.115 & - & 6.67 & 5.03 &   $-$9.7  & 30 & L & 1  \\ 
16 & ATHDFS\_J223253.0$-$603539 &  22:32:53.09  & 0.31 &  $-$60:35:39.7  & 0.31 & 0.090 & - & - & - & 0 & 0 & 0 & 9.2 & L &  2  \\ 
17 & ATHDFS\_J223245.3$-$603537 &  22:32:45.34  & 0.28 &  $-$60:35:37.8  & 0.53 & 0.051 & - & - & - & 0 & 0 & 0 & 5.3 & L &  2  \\ 
18 & ATHDFS\_J223224.0$-$603537 &  22:32:24.02  & 0.03 &  $-$60:35:37.7  & 0.04 & 1.259 & 0.836 & 0.396 & - & 3.53 & 2.27 &  $-$32.3  & 97.1 & L & 1  \\ 
19 & ATHDFS\_J223338.8$-$603523 &  22:33:38.84  & 0.13 &  $-$60:35:23.8  & 0.19 & 0.185 & 0.145 & - & - & 0 & 0 & 0 & 18.5 & L & 1  \\ 
20 & ATHDFS\_J223223.7$-$603520 &  22:32:23.70  & 0.50 &  $-$60:35:20.5  & 0.91 & 0.142 & - & - & - & 9.1 & 4.98 &   $-$4.4  & 6.4 & L &  2  \\ 
21 & ATHDFS\_J223230.2$-$603503 &  22:32:30.21  & 0.13 &  $-$60:35:03.5  & 0.29 & 0.113 & 0.128 & 0.076 & - & 0 & 0 & 0 & 11.9 & C & 2 \\ 
22 & ATHDFS\_J223229.2$-$603459 &  22:32:29.21  & 0.14 &  $-$60:34:59.5  & 0.15 & 0.192 & 0.081 & - & - & 0 & 0 & 0 & 20.3 & L & 1  \\ 
23 & ATHDFS\_J223247.4$-$603450 &  22:32:47.41  & 0.56 &  $-$60:34:50.0  & 0.39 & - & 0.059 & - & - & 0 & 0 & 0 & 5.3 & S &  2 \\ 
24 & ATHDFS\_J223307.1$-$603448 &  22:33:07.18  & 0.55 &  $-$60:34:48.8  & 1.08 & 0.103 & - & - & - & 8.75 & 1.85 & 20.7 & 5.3 & L &  2  \\ 
25 & ATHDFS\_J223212.3$-$603448 &  22:32:12.39  & 0.54 &  $-$60:34:48.7  & 0.49 & 0.093 & - & - & - & 5.21 & 2.35 &  $-$72.7  & 6.9 & L &  2  \\ 
26 & ATHDFS\_J223243.3$-$603442 &  22:32:43.32  & 0.23 &  $-$60:34:42.5  & 0.30 & 0.063 & - & 0.047 & - & 0 & 0 & 0 & 5.1 & C &  2  \\ 
27 & ATHDFS\_J223216.6$-$603434 &  22:32:16.63  & 0.55 &  $-$60:34:34.9  & 0.84 & 0.052 & - & - & - & 0 & 0 & 0 & 5.5 & L &  2  \\ 
28 & ATHDFS\_J223231.6$-$603423 &  22:32:31.69  & 0.26 &  $-$60:34:23.1  & 0.39 & 0.070 & - & - & - & 0 & 0 & 0 & 7.2 & L &  2  \\ 
29 & ATHDFS\_J223245.5$-$603419 &  22:32:45.51  & 0.13 &  $-$60:34:19.0   & 0.16 & 0.265 & 0.150 & 0.06 & - & 0 & 0 & 0 & 6.9 & C & 2 \\ 
30 & ATHDFS\_J223311.9$-$603417 &  22:33:11.97  & 0.41 &  $-$60:34:17.1  & 0.52 & 0.059 & - & - & - & 0 & 0 & 0 & 5.1 & L &  2  \\ 
31 & ATHDFS\_J223327.6$-$603414 &  22:33:27.67  & 0.05 &  $-$60:34:14.3  & 0.07 & 0.456 & 0.498 & 0.262 & - & 1.38 & 0.75 & $-$69.7 & 34.1 & S & 1  \\ 
32 & ATHDFS\_J223329.7$-$603351 &  22:33:29.74  & 0.19 &  $-$60:33:51.9  & 0.26 & 0.152 & 0.081 & 0.073 & - & 0 & 0 & 0 & 5.3 & C &  2  \\ 
33 & ATHDFS\_J223243.4$-$603351 &  22:32:43.47  & 0.06 &  $-$60:33:51.0  & 0.20 & 0.098 & - & 0.066 & 0.114 & 0 & 0 & 0 & 8 & X &  2  \\ 
34 & ATHDFS\_J223306.0$-$603350 &  22:33:06.07  & 0.09 &  $-$60:33:50.3  & 0.13 & 0.452 & 0.288 & 0.123 & - & 4.45 & 2.76 & 11.5 & 31.2 & L & 1  \\ 
35 & ATHDFS\_J223258.5$-$603346 &  22:32:58.59  & 0.02 &  $-$60:33:46.7  & 0.03 & 1.010 & 0.658 & 0.404 & 0.238 & 2.23 & 1.33 & -16.3 & 103.9 & L & 1 \\ 
36 & ATHDFS\_J223225.0$-$603338 &  22:32:25.05  & 0.28 &  $-$60:33:38.7  & 0.64 & 0.080 & - & - & - & 0 & 0 & 0 & 7.5 & L &  2  \\ 
37 & ATHDFS\_J223247.6$-$603337 &  22:32:47.65  & 0.42 &  $-$60:33:37.0  & 0.56 & 0.075 & - & - & - & 0 & 0 & 0 & 6.8 & L &  2  \\ 
38 & ATHDFS\_J223253.1$-$603329 &  22:32:53.15  & 0.58 &  $-$60:33:29.1  & 0.71 & 0.113 & - & - & - & 6.36 & 4.96 &  $-$34.4  & 6.3 & L &  2  \\ 
39 & ATHDFS\_J223337.5$-$603329 &  22:33:37.57  & 0.03 &  $-$60:33:29.1  & 0.04 & 1.126 & 0.718 & 0.373 & - & 2.58 & 1.22 & 9.3 & 88.1 & L & 1  \\ 
40 & ATHDFS\_J223302.8$-$603323 &  22:33:02.83  & 0.61 &  $-$60:33:23.8  & 0.40 & 0.051 & - & - & - & 0 & 0 & 0 & 5.2 & L &  2  \\ 
41 & ATHDFS\_J223306.2$-$603307 &  22:33:06.26  & 0.09 &  $-$60:33:07.9  & 0.35 & - & - & - & 0.071 & 0 & 0 & 0 & 5.6 & X & 2 \\ 
42 & ATHDFS\_J223339.4$-$603306 &  22:33:39.41  & 0.62 &  $-$60:33:06.0  & 0.32 & 0.058 & - & - & - & 0 & 0 & 0 & 5.6 & L &  2  \\ 
43 & ATHDFS\_J223327.9$-$603304 &  22:33:27.95  & 0.13 &  $-$60:33:04.8  & 0.17 & 0.221 & 0.187 & 0.093 & - & 0 & 0 & 0 & 19.8 & L & 1  \\ 
44 & ATHDFS\_J223329.2$-$603302 &  22:33:29.28  & 0.44 &  $-$60:33:02.0  & 0.59 & 0.059 & - & - & - & 0 & 0 & 0 & 5.4 & L & 1  \\ 
45 & ATHDFS\_J223242.6$-$603258 &  22:32:42.66  & 0.06 &  $-$60:32:58.1  & 0.20 & - & - & - & 0.077 & 0 & 0 & 0 & 5.3 & X & 2 \\ 
46 & ATHDFS\_J223234.2$-$603257 &  22:32:34.27  & 0.61 &  $-$60:32:57.4  & 0.79 & 0.057 & - & - & - & 0 & 0 & 0 & 6.1 & L &  2  \\ 
47 & ATHDFS\_J223209.7$-$603253 &  22:32:09.71  & 0.58 &  $-$60:32:53.9  & 0.56 & 0.061 & - & - & - & 0 & 0 & 0 & 5.1 & L &  2  \\ 
48 & ATHDFS\_J223308.6$-$603251 &  22:33:08.60  & 0.06 &  $-$60:32:51.7  & 0.08 & 0.821 & 0.394 & 0.196 & 0.092 & 6.59 & 1.82 &  $-$32.6  & 55 & L & 1  \\ 
49 & ATHDFS\_J223323.2$-$603249 &  22:33:23.25  & 0.03 &  $-$60:32:49.2  & 0.04 & 0.457 & 0.561 & 0.360 & 0.223 & 0 & 0 & 0 & 32.9 & C & 1  \\ 
50 & ATHDFS\_J223229.5$-$603243 &  22:32:29.55  & 0.07 &  $-$60:32:43.6  & 0.16 & 0.237 & 0.140 & 0.108 & - & 0 & 0 & 0 & 10.5 & C & 1  \\ 
51 & ATHDFS\_J223212.9$-$603243 &  22:32:12.95  & 0.04 &  $-$60:32:43.3  & 0.06 & 1.466 & 0.903 & 0.657 & - & 5.84 & 4.24 &   $-$5.5  & 77.9 & L & 1  \\ 
52 & ATHDFS\_J223317.7$-$603235 &  22:33:17.75  & 0.41 &  $-$60:32:35.2  & 0.38 & 0.070 & - & - & - & 0 & 0 & 0 & 7.3 & L &  2  \\ 
53 & ATHDFS\_J223212.9$-$603234 &  22:32:12.90  & 0.03 &  $-$60:32:34.6  & 0.03 & 2.816 & 1.574 & 0.716 & - & 7.07 & 2.77 &  $-$54.7  & 147.3 & L & 1  \\ 
54 & ATHDFS\_J223335.3$-$603234 &  22:33:35.31  & 0.35 &  $-$60:32:34.5  & 0.55 & 0.054 & - & - & - & 0 & 0 & 0 & 5.5 & L &  2  \\ 
55 & ATHDFS\_J223331.6$-$603222 &  22:33:31.64  & 0.08 &  $-$60:32:22.2  & 0.11 & 0.395 & 0.286 & 0.120 & - & 3.31 & 2.46 & 2.4 & 32.4 & L & 1  \\ 
56 & ATHDFS\_J223302.1$-$603213 &  22:33:02.18  & 0.33 &  $-$60:32:13.2  & 0.43 & 0.063 & - & - & - & 0 & 0 & 0 & 6.4 & L &  2  \\ 
57 & ATHDFS\_J223303.1$-$603132 &  22:33:03.17  & 0.35 &  $-$60:31:32.8  & 0.47 & 0.052 & - & - & - & 0 & 0 & 0 & 5.2 & L &  2  \\ 
58 & ATHDFS\_J223254.4$-$603131 &  22:32:54.41  & 0.47 &  $-$60:31:31.5  & 0.41 & 0.139 & - & - & - & 6.91 & 3.17 & 64.1 & 9 & L &  2  \\ 
59 & ATHDFS\_J223316.0$-$603127 &  22:33:16.09  & 0.43 &  $-$60:31:27.6  & 0.67 & 0.052 & - & - & - & 0 & 0 & 0 & 5.3 & L &  2  \\ 
60 & ATHDFS\_J223256.4$-$603058 &  22:32:56.42  & 0.39 &  $-$60:30:58.9  & 0.35 & 0.074 & - & - & - & 0 & 0 & 0 & 8.5 & L &  2  \\ 
61 & ATHDFS\_J223304.8$-$603031 &  22:33:04.89  & 0.44 &  $-$60:30:31.2  & 0.25 & 0.084 & - & - & - & 0 & 0 & 0 & 8.2 & L &  2  \\ 
62 & ATHDFS\_J223241.4$-$603025 &  22:32:41.42  & 0.30 &  $-$60:30:25.9  & 0.27 & 0.092 & - & - & - & 0 & 0 & 0 & 9.4 & L &  2  \\ 
63 & ATHDFS\_J223216.6$-$603016 &  22:32:16.67  & 0.54 &  $-$60:30:16.7  & 0.51 & 0.052 & - & - & - & 0 & 0 & 0 & 5.2 & L &  2  \\ 
64 & ATHDFS\_J223303.9$-$603013 &  22:33:03.96  & 0.46 &  $-$60:30:13.6  & 1.05 & 0.052 & - & - & - & 0 & 0 & 0 & 5.1 & L &  2  \\ 
65 & ATHDFS\_J223331.1$-$603007 &  22:33:31.14  & 0.31 &  $-$60:30:07.9  & 0.40 & 0.094 & - & - & - & 0 & 0 & 0 & 9 & L &  2  \\ 
66 & ATHDFS\_J223224.7$-$603005 &  22:32:24.77  & 0.66 &  $-$60:30:05.5  & 0.47 & 0.055 & - & - & - & 0 & 0 & 0 & 5.1 & L &  2  \\ 
67 & ATHDFS\_J223236.5$-$603000 &  22:32:36.56  & 0.02 &  $-$60:30:00.6  & 0.03 & 1.507 & 0.877 & 0.394 & - & 3.71 & 1.33 &  $-$30.0  & 129.7 & L & 1  \\ 
68 & ATHDFS\_J223253.7$-$602946 &  22:32:53.78  & 0.38 &  $-$60:29:46.7  & 0.52 & 0.045 & - & - & - & 0 & 0 & 0 & 5.1 & L &  2  \\ 
69 & ATHDFS\_J223316.8$-$602934 &  22:33:16.80  & 0.02 &  $-$60:29:34.7  & 0.05 & 1.003 & 0.811 & 0.569 & - & 1.45 & 0.44 & 1.5 & 37.4 & C & 1  \\ 
70 & ATHDFS\_J223329.1$-$602933 &  22:33:29.16  & 0.12 &  $-$60:29:33.4  & 0.22 & 0.261 & - & - & - & 5.36 & 1.3 & 3.5 & 20.3 & L & 1  \\ 
71 & ATHDFS\_J223301.8$-$602930 &  22:33:01.82  & 0.41 &  $-$60:29:30.2  & 0.72 & 0.071 & - & - & - & 0 & 0 & 0 & 7.1 & L & 1  \\ 
72 & ATHDFS\_J223303.0$-$602927 &  22:33:03.00  & 0.34 &  $-$60:29:27.1  & 0.67 & 0.116 & - & - & - & 7.22 & 3.83 &   $-$1.4  & 8.2 & L & 1  \\ 
73 & ATHDFS\_J223317.7$-$602916 &  22:33:17.72  & 0.50 &  $-$60:29:16.0  & 0.72 & 0.059 & - & - & - & 0 & 0 & 0 & 5.2 & L &  2  \\ 
74 & ATHDFS\_J223236.2$-$602855 &  22:32:36.20  & 0.31 &  $-$60:28:55.6  & 0.62 & 0.079 & - & - & - & 0 & 0 & 0 & 7.5 & L &  2  \\ 
75 & ATHDFS\_J223307.7$-$602853 &  22:33:07.73  & 0.25 &  $-$60:28:53.8  & 0.55 & 0.068 & - & - & - & 0 & 0 & 0 & 6.5 & L &  2  \\ 
76 & ATHDFS\_J223326.9$-$602850 &  22:33:26.94  & 0.15 &  $-$60:28:50.0  & 0.22 & 0.182 & - & - & - & 2.93 & 1.13 &   $-$5.7  & 16.3 & L & 1  \\ 
77 & ATHDFS\_J223330.5$-$602849 &  22:33:30.55  & 0.51 &  $-$60:28:49.7  & 0.90 & 0.047 & - & - & - & 0 & 0 & 0 & 5 & L &  2  \\ 
78 & ATHDFS\_J223307.0$-$602827 &  22:33:07.07  & 0.11 &  $-$60:28:27.8  & 0.15 & 0.354 & 0.192 & 0.101 & - & 4.58 & 2.6 & 32.4 & 26.6 & L & 1  \\ 
79 & ATHDFS\_J223255.9$-$602810 &  22:32:55.99  & 0.20 &  $-$60:28:10.1  & 0.38 & 0.108 & 0.244 & - & - & 0 & 0 & 0 & 10.3 & L &  2  \\ 
80 & ATHDFS\_J223240.7$-$602755 &  22:32:40.71  & 0.34 &  $-$60:27:55.3  & 0.38 & 0.139 & - & - & - & 3.77 & 2.56 & 68 & 9.2 & L &  2  \\ 
81 & ATHDFS\_J223311.5$-$602725 &  22:33:11.53  & 0.32 &  $-$60:27:25.1  & 0.40 & 0.344 & - & - & - & 9.06 & 6.09 & 33.7 & 13.2 & L &  2  \\ 
82 & ATHDFS\_J223241.5$-$602719 &  22:32:41.50  & 0.09 &  $-$60:27:19.8  & 0.13 & 0.351 & - & - & - & 0 & 0 & 0 & 28 & L & 1  \\ 
83 & ATHDFS\_J223244.5$-$602719 &  22:32:44.57  & 0.33 &  $-$60:27:19.2  & 0.60 & 0.088 & - & - & - & 0 & 0 & 0 & 7 & L &  2  \\ 
84 & ATHDFS\_J223317.1$-$602714 &  22:33:17.11  & 0.38 &  $-$60:27:14.1  & 0.73 & 0.122 & - & - & - & 8.19 & 0.86 &  $-$22.1  & 7.6 & L &  2  \\ 
85 & ATHDFS\_J223312.3$-$602707 &  22:33:12.30  & 0.41 &  $-$60:27:07.9  & 0.51 & 0.078 & - & - & - & 0 & 0 & 0 & 7 & L &  2  \\ 
86 & ATHDFS\_J223257.4$-$602657 &  22:32:57.44  & 0.39 &  $-$60:26:57.4  & 0.49 & 0.083 & - & - & - & 0 & 0 & 0 & 7.3 & L &  2  \\ 
87 & ATHDFS\_J223259.9$-$602654 &  22:32:59.93  & 0.33 &  $-$60:26:54.8  & 0.69 & 0.061 & - & - & - & 0 & 0 & 0 & 5.5 & L &  2  \\ 
\enddata
\end{deluxetable}

\begin{deluxetable}{llrrrrrrrrrrrl}

\tabletypesize{\scriptsize}
\rotate
\tablewidth{0pt}
\tablecaption{ Optical and derived properties of the subset}
\label{discuss3}
\tablehead{
\colhead {Ref } & 

\colhead{Name } & 
\colhead{$S_{20}$} & 
\colhead{z } & 
\colhead{V } & 
\colhead{R } & 
\colhead{I } & 
\colhead{J } & 
\colhead{K } & 
\colhead{ log ($L_{20}$) } & 
\colhead{log ($S_{20}/I)$ } & 
\colhead{SFR } & 
\colhead{Classification } & 
\colhead{Notes }
\\
\colhead {} & 
\colhead{} & 
\colhead{(mJy)} & 
\colhead{} & 
\colhead{} & 
\colhead{} & 
\colhead{} & 
\colhead{} & 
\colhead{} & 
\colhead{$WHz^{-1}$ } & 
\colhead{} & 
\colhead{$M_{0}yr^{-1}$} & 
\colhead{} & 
\colhead{}
}
\startdata
3 & ATHDFS\_J223245.6$-$603857 & 0.843 & 0.75 & 25.45 & 24.6 & 23.39 &
&  & 23.63 & 2.72 & 515 & SB? & \\
7 & ATHDFS\_J223254.5$-$603748 & 0.092 & 0.1798 & 19.11 & 18.6 & 17.91 &
&  & 21.69 & $-$0.43 & 6 & Sy & \\
12 & ATHDFS\_J223316.5$-$603627 & 0.649 & 0.29 & 21.75 & 21.2 & 20.68 &
&  & 22.89 & 1.52 & 95 & spiral SB& \\
19 & ATHDFS\_J223338.8$-$603523 & 0.185 & 0.2250 & 20.04 & 19.3 & 18.61
& 18.28 & 17.74 & 22.17 & 0.15 & 19 & elliptical AGN & b \\
26 & ATHDFS\_J223243.3$-$603442 & 0.063 & 0.4233 & 20.55 & 19.6 & 18.87
& 18.72 & 18.09 & 22.15 & $-$0.21 & 17 & spiral AGN? & b \\
29 & ATHDFS\_J223245.5$-$603419 & 0.265 & 0.4603 & 21.95 & 21.1 & 20.33
& 19.84 & 18.94 & 22.83 & 1.00 & 81 & SB  & a\\
31 & ATHDFS\_J223327.6$-$603414 & 0.456 & 0 & 0 & $>$28 & $>$26 &  &   &
& $>$3.50 &  & GPS & h\\
33 & ATHDFS\_J223243.4$-$603351 & 0.098 & 1.5660 & 20.28 & 19.9 & 19.45
& 19.78 & 19.53 & 23.07 & 0.21 & 142 & QSO & a \\
34 & ATHDFS\_J223306.0$-$603350 & 0.452 & 0.1733 & 17.77 & 17.2 & 16.58
& 16.68 & 16.36 & 22.36 & $-$0.27 & 27 & SB & a \\
35 & ATHDFS\_J223258.5$-$603346 & 1.010 & 1.69 & 0 & $>$25 & 25.75 &
23.44 & 22.31 & 24.12 & 3.74 & 1579 & Dusty AGN? & d \\
37 & ATHDFS\_J223247.6$-$603337 & 0.075 & 0.5803 & 21.05 & 20.0 & 19.16 &
18.86 & 18.17 & 22.43 & $-$0.02 & 32 & Massive spiral & a \\
39 & ATHDFS\_J223337.5$-$603329 & 1.126 & 2.2380 & 17.16 & 16.9 & 16.51
& 16.92 & 16.40 & 24.28 & 0.10 & 2280 & RL QSO & b \\
43 & ATHDFS\_J223327.9$-$603304 & 0.221 & 0 & $>$25 & $>$25 & $>$23.5 &
23.91 & 22.42 &  & $>$3.19 &  & obscured AGN & b \\
48 & ATHDFS\_J223308.6$-$603251 & 0.821 & 0.64 & 23.07 & 21.8 & 20.45 &
&  & 23.53 & 1.53 & 405 & AGN & \\
49 & ATHDFS\_J223323.2$-$603249 & 0.457 & 0 & 0 & $>$25 & 0 & 25.05 &
24.34 &   & 3.17 &  & GPS & b,f \\
55 & ATHDFS\_J223331.6$-$603222 & 0.395 & 0.4652 & 21.24 & 20.6 & 19.9 &
19.58 & 18.78 & 23.01 & 1.00 & 123 & obscured SB & b \\
67 & ATHDFS\_J223236.5$-$603000 & 1.507 & 0 & $>$25 & $>$25 & $>$23.5 &
&  & & $>$3.02 &  & ? &\\

69 & ATHDFS\_J223316.8$-$602934 & 1.003 & 0.12 & 22.53 & 22.7 & 21.05 &
&  & 22.42 & 1.86 & 35 & AGN & \\

70 & ATHDFS\_J223329.1$-$602933 & 0.261 & 0 & 0 & 23.9 & 0 &  &  & &
2.09 &  & AGN & g\\
\enddata         

\tablenotetext{a}{J and K from Francheschini et al, 2003. }
\tablenotetext{b}{J and K from EIS. }
\tablenotetext{c}{J has large uncertainty. }
\tablenotetext{d}{I,J,K from Vanzella et al 2001.}
\tablenotetext{e}{For the calculation of the radio-optical ratio, the 20
cm flux density is assumed to be equal to the 3cm flux density.}
\tablenotetext{f}{For the calculation of the radio-optical ratio,  I has
been estimated as I=J+0.12, where 0.12 is the mean value of I-J for
those sources in this subset for which a measured value of I-J is
available.}
\tablenotetext{g}{For the calculation of the radio-optical ratio,  I has
been estimated as I=R$-$0.83, where 0.83 is the mean value of R-I for
those sources in this subset for which a measured value of R-I is
available.}
\tablenotetext{h}{R and I limits based on non-detection in WFPC flanking
field observations.}

\end{deluxetable}
\clearpage

\begin{figure}
\plotone{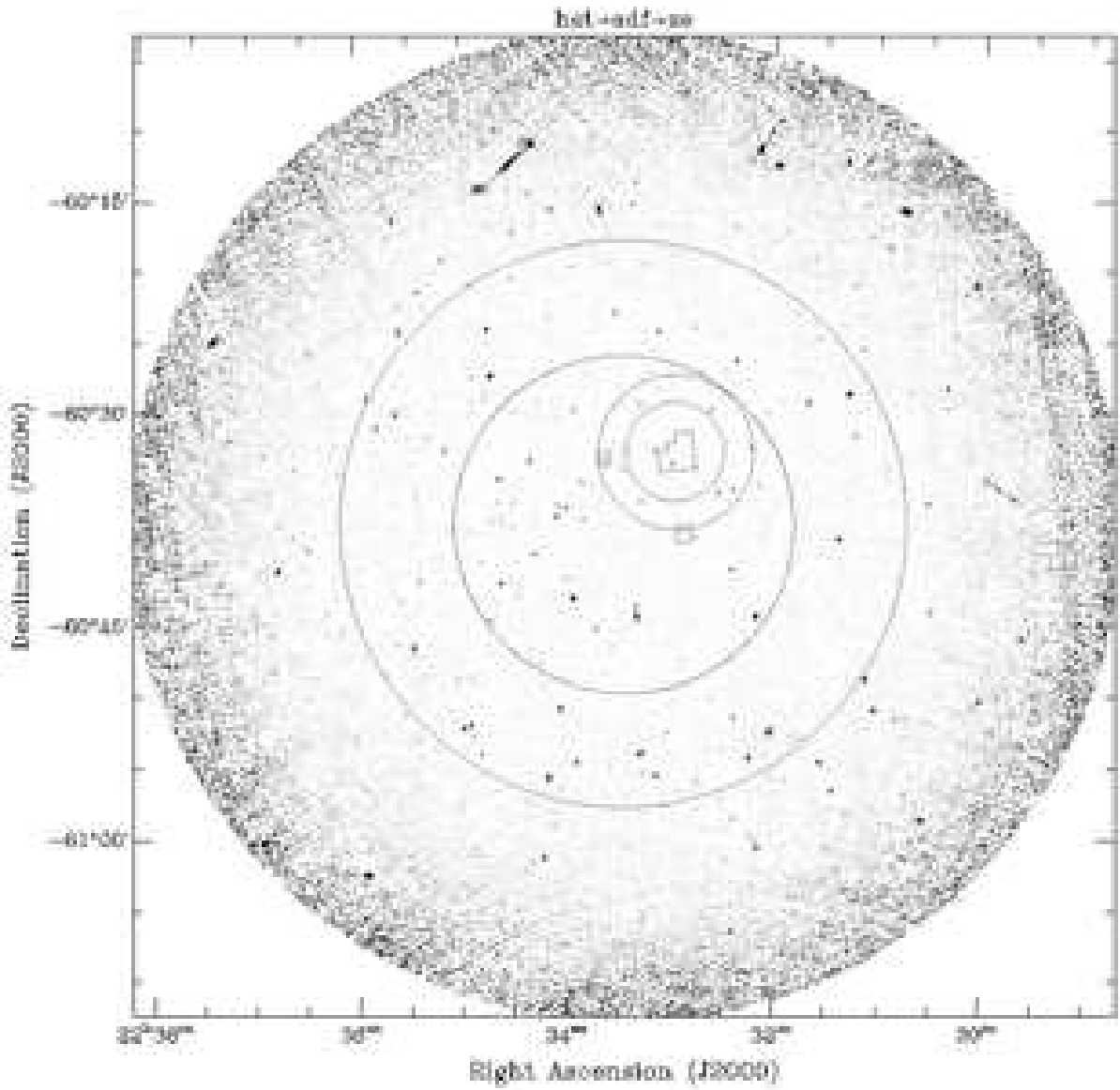}
\caption{Position of the 4 pointings of the ATHDFS relative
to the HST WFPC2 field. The background image is the size of the entire \faa\
primary beam. Working inwards to the WFPC2 field, the 
inner circles represent the catalogued areas of
the 20, 11, 6, and 3 cm primary beams respectively, which are set to the level where the sensitivity falls to 39\% of that at the beam centre. The three polygons show the positions of the HST WFPC field (top right), NICMOS field (lower), and STIS QSO field (left). }
\end{figure}

\begin{figure}
\plotone{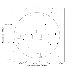}
\caption{ATCA \faa\ image of the region containing the sample of 87 sources listed in Table 3. This sample includes all sources detected, at any wavelength, within the circle, which has a radius of 6.5 arcmin.}
\end{figure}

\clearpage 

\begin{figure}[3]

\includegraphics[scale=0.5]{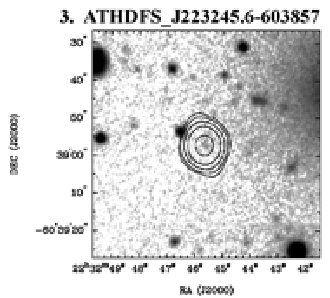}
\includegraphics[scale=0.5]{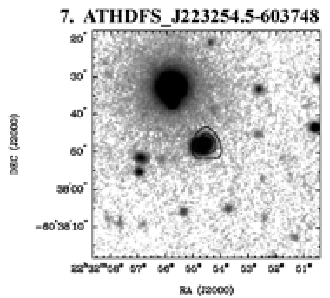}
\includegraphics[scale=0.5]{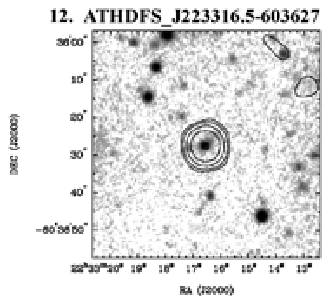}
\includegraphics[scale=0.5]{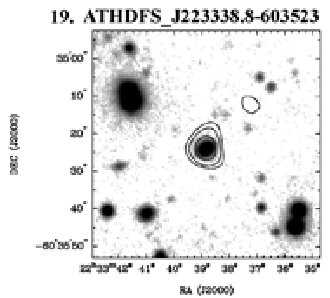}
\includegraphics[scale=0.5]{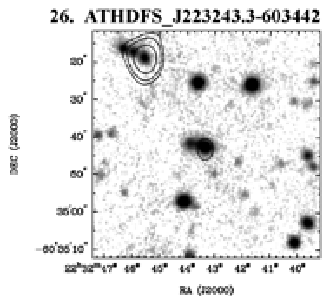}
\includegraphics[scale=0.5]{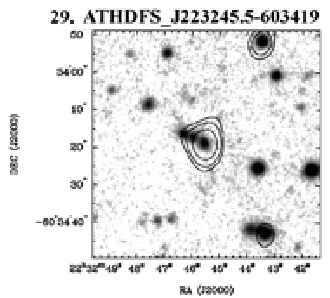}
\end{figure}

\begin{figure}[3]
\includegraphics[scale=0.5]{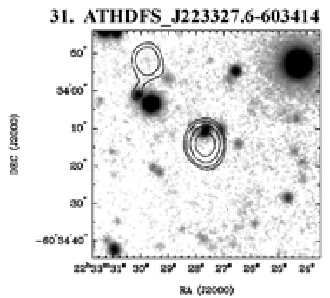}
\includegraphics[scale=0.5]{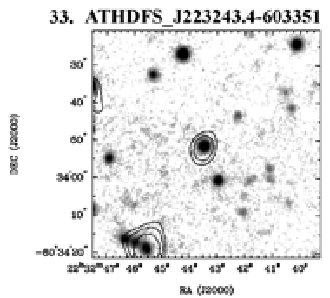}
\includegraphics[scale=0.5]{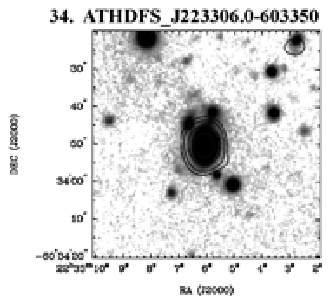}
\includegraphics[scale=0.5]{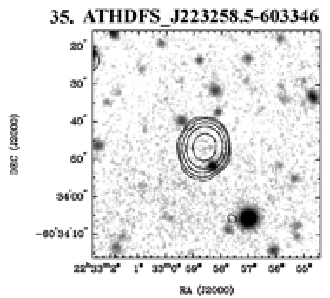}
\includegraphics[scale=0.5]{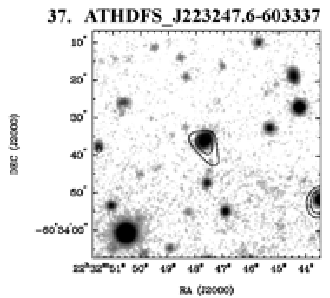}
\includegraphics[scale=0.5]{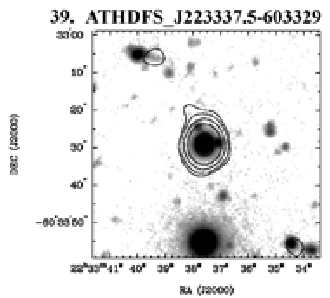}
\end{figure}

\begin{figure}[3]
\includegraphics[scale=0.5]{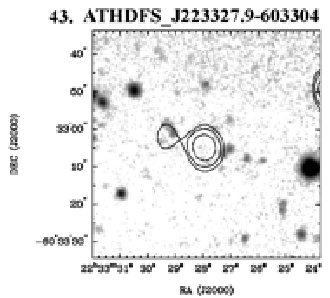}
\includegraphics[scale=0.5]{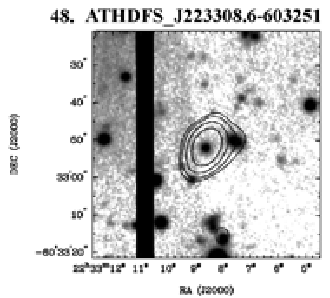}
\includegraphics[scale=0.5]{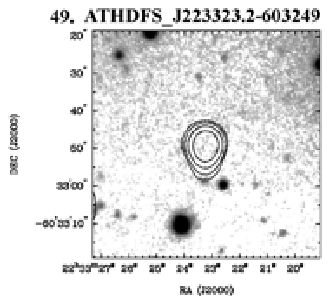}
\includegraphics[scale=0.5]{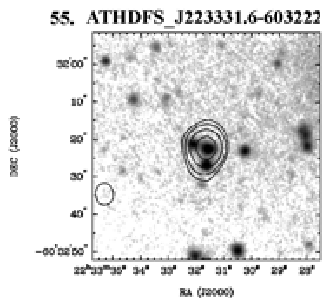}
\includegraphics[scale=0.5]{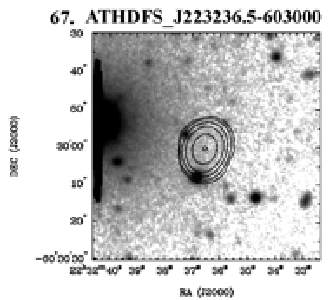}
\end{figure}

\begin{figure}[3]
\includegraphics[scale=0.5]{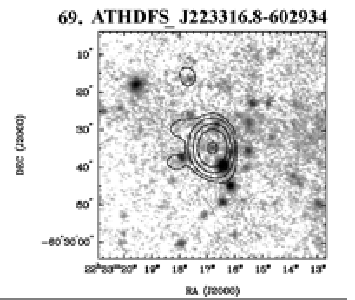}
\includegraphics[scale=0.5]{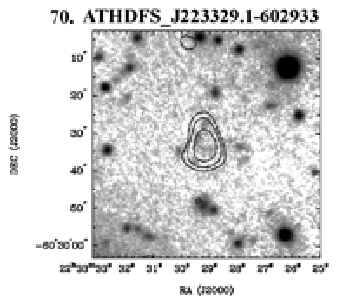}

\caption{Radio images (contours) overlaid on CTIO I-band images
(greyscale) for each of the sources in our subset. In each case, the named source is at the centre of the image. The radio image of source 41 (ATHDFS\_2J223306.2$-$603307) shows 3 cm data; all others show 20 cm data. Contours are at 3, 5, 10, 20, 50 and 100 $\times$ the local rms.}

\end{figure}

\end{document}